# Crowdsourcing Felt Reports using the MyShake smartphone app

Authors: Qingkai Kong, Richard M Allen, Steve Allen, Theron Bair, Akie Meja, Sarina Patel, Jennifer Strauss, Stephen Thompson.

Corresponding author: Qingkai Kong

Address: 7000 East Ave, Livermore, CA 94550

Email: kongqk@berkeley.edu

Abstract:

MyShake is a free citizen science smartphone app that provides a range of features related to earthquakes. Features available globally include rapid post-earthquake notifications, live maps of earthquake damage as reported by MyShake users, safety tips and various educational features. The app also uses the accelerometer in the mobile device to detect earthquake shaking and to record and submit waveforms to a central archive. In addition, MyShake delivers earthquake early warning alerts in California, Oregon and Washington. In this study we compare the felt shaking reports provided by MyShake users in California with the U.S. Geological Survey's "Did You Feel It?" intensity reports. The MyShake app simply asks, "What strength of shaking did you feel?" and users report on a five-level scale. When the MyShake reports are averaged in spatial or time bins, we find strong correlation with the Modified Mercalli Intensity scale values reported by the USGS based on the DYFI surveys. The MyShake felt reports can therefore contribute to the creation of shaking intensity maps.

**Introduction**

"How strong was the shaking?" is a question we usually ask after each earthquake. From the very earliest written records in our human history, to modern seismological instrumentation, we have tried to provide more quantitative answer to this question. More recently, with the wide adoption of the internet, computers and smartphones, new crowdsourcing methods to answer this question have been developed. These methods include surveys to provide felt reports (Wald *et al.*, 2001; Wald and Dewey, 2005; Atkinson and Wald, 2007; Bossu *et al.*, 2012, 2015, 2018; Rochford *et al.*, 2018; Liang *et al.*, 2019; Quitoriano and Wald, 2020), using messages from Twitter (Earle, 2010; Earle *et al.*, 2010; Sakaki *et al.*, 2010; Ruan *et al.*, 2020, 2022), and smartphones or standalone low-cost sensors (Cochran *et al.*, 2009; Clayton *et al.*, 2012, 2015; Hsieh *et al.*, 2014; Minson *et al.*, 2015; Wu, 2015; Kong *et al.*, 2016; Jan *et al.*, 2018; Nof *et al.*, 2019; Steed *et al.*, 2019). While the high-quality research grade regional seismic and geodetic networks provide precise but sparse observations, these crowdsourcing approaches provide a dense but noisier view of earthquake shaking. They also provide information about people's perception of shaking and observations of damage after an earthquake.

Within these newly developed crowdsourcing approaches, MyShake is an application developed for smartphones at the Berkeley Seismology Lab. It utilizes both the sensors inside smartphones and user-uploaded felt reports after an earthquake to learn more about the distribution of shaking and its impacts (Allen *et al.*, 2019; Strauss *et al.*, 2020). To learn more about how MyShake uses the sensors inside the phones to detect earthquakes for earthquake early warning, please refer to (Kong *et al.*, 2020). In this paper, we will focus on the MyShake users' felt reports, provided through a short series of questions that the users can complete to evaluate the shaking after an earthquake. In particular, we compare these MyShake felt reports to the U.S. Geological Survey's Did You Feel It (DYFI) observations representing the "gold standard" for such observations. We show examples of intensity maps derived from the MyShake reports and the distribution of responses. Furthermore, we show the strong correlation between the MyShake felt reports and those from the DYFI system. Even though the MyShake questions are very simple compared to the DYFI survey, with sufficient reports from a large group of the users in the earthquake region, these measurements can be useful and strongly correlate to the known intensity scale with careful calibration.

**Overview of the MyShake felt report**

The MyShake felt report is a set of four questions for the users to answer to provide information about their experiences during the earthquake and observations of damage around them after an earthquake (Rochford *et al.*, 2018). With both simplicity and usefulness in mind, the felt report questionnaires are designed to minimize the effort for the user to complete a report after an earthquake. There are a total of 4 questions asked in the questionnaires as shown in Table 1, where questions 2 - 4 are aided with pictorial representations to assist the user. An example of question 2 with selection options is shown in Figure 1. The user only needs to scroll through the images/descriptions to a level that matches his/her experience. The felt report completed by a user is then uploaded to the MyShake server. Each report immediately becomes part of an aggregated map of felt shaking intensity visible to all MyShake users in the app. The map provides an immediate visualization of the strength of shaking for the area and a user can click on a location to see the number of reports and the different levels of shaking and damage. See figure 3 in (Strauss *et al.*, 2020) for an example of reported shaking maps with information for a recent earthquake in Puerto Rico.

In this study, we focus on earthquakes that have felt reports within California between 2019-10-15 to 2021-05-11 (based on the ANSS Comprehensive Earthquake Catalog from USGS). In each felt report, we get the timestamp when the report arrived at the server, the location of the user making the submission, and the shaking scale. See table 2 for an example. The shaking scale is an integer number, with -1, 0, 1, 2, 3, representing none, light shaking, moderate shaking, strong shaking, and severe shaking, respectively. In order to compare and calibrate the shaking scales in the MyShake felt reports, we obtained the corresponding DYFI data from the USGS. Specifically, we downloaded Intensity vs. Distance, Responses vs. Time, and the Intensity Map from USGS earthquake websites.

**Comparison with Did You Feel It Intensity**

Each USGS DYFI survey response is converted into an estimate of seismic intensity on the Modified Mercalli Intensity (MMI) scale (Wald *et al.*, 2011). Here, we will compare the MyShake reported shaking intensity scale with 5 levels, to the 10 level DYFI MMI. Our goal is to determine the degree to which the simplified MyShake survey provides intensity estimates similar to the DYFI survey. If they do correlate, then we want to develop a scaling relation that will allow us to convert the MyShake intensity shaking scale to MMI. The MyShake felt report system is still quite new, and for most earthquakes provides far fewer felt reports than the USGS DYFI submissions. Still, a good number of events have more than a few thousands reports submitted, which enables us to take the first step to understand what these reports can tell us.

Figure 2 shows a histogram of the number of reports collected for each event from the period 2019-10-15 to 2021-05-11 in California. In total, there are 325 events that have at least 1 felt report, and the majority of the events have less than 500 reports, with 29 events having more than 500 reports and 20 events having more than 700 reports for each earthquake. Table 3 lists these 20 events with the number of felt reports that we later use for developing a conversion between MyShake intensity shaking scale to the DYFI MMI intensity. A relationship between the number of felt reports and the population within 0.5 degrees of the event is shown in figure 3, with colors representing the magnitude. The population data are extracted from the 2020 Gridded Population of the World V4 (see data resources section). We can see the general trend: the number of reports increases when larger populations are nearby and for larger magnitude earthquakes.

We use 16 of the 20 events that have more than 700 felt reports to build a linear relationship between the MyShake felt report shaking scale (-1, 0, 1, 2, 3) and the MMI intensity scale from DYFI data. We reserve the rest 4 events shown in bold in table 3 as test set. An example of the raw MyShake intensities from the felt reports is shown in figure S1. We first bin the MyShake and DYFI intensity observations as a function of hypocentral distance. In order to calibrate with the USGS DYFI intensities, we use the same distance bins as that used in the USGS Intensity vs. Distance plots up to a maximum distance of 308 km. The distance bins are 5.5, 7.3, 9.7, 13.0, 17.3, 23.1, 30.8, 41.1, 54.8, 73, 97.4, 129.9, 173.2, 231.0, and 308.0 km. For MyShake, to ensure a robust shaking estimate, we only compute the average shaking report value when there are 10 or more felt reports within a specific distance bin (see figure 9 in the discussion section supporting this choice). This provides us 120 data points from the 16 events for building the relationship as well as 34 data points for testing. The averaged intensity scales within the distance bins from the MyShake felt reports and USGS DYFI MMIs are then used to determine a simple linear relationship:

$$MMI\_intensity = a + MyShake\_Shaking\_Scale \times b \qquad (1)$$

We use the least square regression, and found a = 2.3 and b = 2.44 yield the best results. This relation maps the -1 (None), 0 (light), 1 (moderate), 2 (strong), 3 (severe) of the MyShake felt report shaking scale to 0.0 (converted to 0 for negative intensities), 2.3, 4.7, 7.2 and 9.6 on the MMI scale respectively. Table 4 summarizes the results of the regression. The p-values associated with the two coefficients based on the t statistical tests are all less than 5%, which indicate they are all statistically significant.

The fact that the MyShake felt reports have only 5 levels, might be interpreted to suggest that they provide less granular information than the 10-level DYFI MMI data. However, once the MyShake data are binned and averaged, it provides very similar information to the DYFI data. Figure 4 shows the four earthquakes in the test set and compares the converted MMI shaking scale from the MyShake data to the DYFI MMI shaking scale. Both datasets are averaged within each hypocentral distance bin. Similar comparisons for a random selection of 10 training events are included in the supplementary material (Figure S2 - S11). From these figures, we can see the majority of the events show good agreement between the two independent shaking estimates when there are enough MyShake felt reports to aggregate. Furthermore, figure 5 shows the fit between the calibrated MyShake felt report intensity and DYFI for the 20 events that have 700 or more reports, i.e. both for training and the testing sets. The Mean Absolute Error (MAE) in figure 5a is calculated by taking the absolute value of the errors in each distance bin, and then taking the average for each event. These values show the overall fit for individual events. The vertical bars show the uncertainties of the MAE values in different distance bins within each event. Overall, more reports available will generally reduce the uncertainties.

Figure 5b shows the scatter plot between the USGS DYFI and MyShake calibrated intensities for each distance bins for all the events, which gives us a global view of the overall fit. 136 sample bins out of 154 have the intensity difference within 0.5 unit, which is 88.3% of the data. The corresponding histogram of the intensity differences from figure 5b is shown in the figure 5c. Both the training and testing set are shown, and the differences are both centered around 0, with a standard deviation around 0.35.

Figure 6 shows the spatial distribution of these 20 events, and the variations in MAE for each event. The background population counts in 5 km grids are also shown. There is a general trend that the events outside the densely populated areas have larger errors likely due to the smaller number of felt reports collected. An interesting exception is the M5.8 Lone Pine event (the one with text label in the figure) which has a very low error relative to the DYFI data. We can see from figure S9 for the Lone Pine event, that even though there is little population close by, it has 3 bins between 100 and 200 km, where each of them has more than 80 felt reports. For these bins, the correspondence between the DYFI intensity and the converted MyShake intensity is very high.

Not only can we derive the intensity vs. distance using MyShake felt reports (as shown in figure 4), we can also generate a map of intensity variations similar to a ShakeMap. Using felt reports from the September 19, 2020, M4.5 earthquake in Los Angeles (ci38695658) as an example, figure 7a shows the spatial distribution of the MyShake calibrated intensities in 10 km UTM boxes (same as USGS DYFI UTM boxes). Figure 7b compares the derived intensities to the corresponding USGS DYFI intensities within each UTM box. The mean residuals is -0.08 unit (MyShake – DYFI) and standard deviation is 0.75. Most of the residuals are small across the region. The figure 7c and 7d show more detailed shaking distribution in 1 km UTM boxes, but due to there are many boxes only have less than 10 reports, the residuals are larger comparing with the 10 km UTM version. From both the 10 and 1 km UTM maps, we see similar shaking patterns comparing to DYFI maps: strong shaking about intensity IV to V around the epicenter, with stronger shaking over a wider region to the west of the earthquake. Figure S13-S15 shows a similar spatial comparison for the 2021-01-17 M4.2 earthquake (nc73512355).

We also plot the submission timeline of the MyShake and DYFI reports for the 2020-09-19 M4.5 earthquake in figure 8 (see another example in figure S16 for event nc73512355). Though the general patterns are similar on both platforms, MyShake submissions are slightly slower in terms of the percentage of the total submission after the first 50%~60% submissions, this may be due to the different behaviors of smartphone and internet users, which needs more observations and analysis to confirm.

**Discussion**

MyShake's felt report system is new and it will take some time for users to adapt and get used to it, both to report damage and also to use the live map in the app to see where damage has occurred in an earthquake. It is important for the success of citizen science projects to provide interactive features that show the utility of a users' engagement. In the case of MyShake, showing the users a community derived shaking map and the number of people who felt an earthquake near them provides a sense of participation and community. So far, with the current density of MyShake users, for M3.5-5.5 earthquakes, we have collected hundreds to a few thousands felt reports. To minimize the required effort by the users, the felt reports were kept very simple, but we did not know at the time of roll-out if this data could still be used to estimate shaking intensity in the same way that the more sophisticated DYFI reports do. The work reported above, suggests that the 5 shaking-level scale in MyShake can indeed be converted to the MMI scale using a simple linear equation. While the relationship developed here provides initial useful shaking information, it has several limitations and needs to be verified and improved in the future as more events are recorded. In particular, there are three possible areas for improvement. First, we note that the current dataset does not include examples of the strongest shaking intensities, i.e. greater than MMI 5. This is key for an accurate shaking estimate for larger earthquakes, which usually draw more attention from the public. Second, the current linear relationship between MyShake and DYIF MMI should be re-analyzed when larger amount of data is available. Finally, the linear relationship that we developed based purely on distance bins may be enhanced by exploring some non-linear conversion relationships.

Since the MyShake felt reports are relying on crowdsourced estimates, we generally expect that having more reports within each binned location would yield more stable results. Figure 9 plots the difference between calibrated MyShake and USGS DYFI data aggregated in 1km UTM boxes versus the number of MyShake felt reports. We can see the standard deviations of the differences decrease with increased number of reports. Between 1 and 5 reports per bin there is a rapid decrease in the errors, with more improvement as the number of reports increases to 10. For this reason, we required 10 reports per bin for the development of the regression relation to make stable estimates. From figure 7, by comparing the 10 km and 1 km UTM boxes, we can see this effect clearly. Since there are more reports to aggregate in the 10 km UTM boxes, the mean and standard deviation of the residuals are smaller comparing with the map of the 1 km UTM boxes.

To illustrate some of the apparent differences/discrepancies between the MyShake and DYFI felt report values, figure 10 shows two events that have larger intensity discrepancies. The M3.8 Morgan Hill event (Figure 9a) shows discrepancies in just a few of the distance bins, all of which have smaller numbers of reports. Even with the requirement of 10 reports per bin, there may still be some large anomalies in bins at the lower end of the number of reports. In the future, we may consider ways to remove large anomaly values. Figure 9b shows the M4.7 event near Truckee for which we can see the MyShake converted intensities are systematically higher than that from DYFI, by about 0.5 - 0.7 units. Most of the distance bins (two exceptions) have 50 or more felt reports. We do not have an explanation for this event with systematically different intensities. One possibility might be that people in different regions may have different interpretations of the 5 levels of shaking as described in the MyShake tool. For example, people in regions with many earthquakes may have more opportunity to calibrate their sense of shaking. However, with a relatively small dataset, it is not obvious why the reports from MyShake users and DYFI users would be different in various locations.

This work is also a step towards the development of a citizen science platform that can utilize multiple data sources to study earthquakes and their impact. The MyShake felt reports are a relatively new feature in the MyShake app with the aim of collecting users' observations about the earthquake after it occurred. This data is complementary to the waveforms recorded with the accelerometer in the phone. The data collected from the felt reports are more subjective and depends on the sensitivity and interpretation of each person, but we show here that by aggregating large numbers of these felt reports in a region, the averaged ground shaking reports are consistent with the more detailed DYFI surveys and reports. Also, even though the MyShake reports are based on a small number of categorical levels, i.e. none, light, moderate, strong, severe shaking, once averaged in spatial bins, the averaged values can provide a more granular estimate of shaking intensity than the in-app 5 report levels. This has the potential to provide a complimentary source of data to the USGS DYFI reports.

Looking forward, the collection of felt reports can be assisted by allowing the users to upload post-earthquake images of damage to buildings, roads, or other infrastructures. These images could serve several purposes, first, by displaying the images in the MyShake felt reports map, they can be used to increase the feeling of user participation. Second, these damage images can be potentially used in the civil engineering community for damage estimation or more detailed understanding of the shaking in the region. Initial work has been developed (Chachra *et al.*, 2022) to use transfer learning to identify damage building images from crowdsourcing platforms, in the hope to filter out unrelated pictures uploaded by the users.

**Conclusion**

This paper shows our initial efforts to evaluate and link the simple MyShake felt reports to MMI shaking as reported by the USGS DYFI product. We find good correlation between the two using a simple linear relationship (equation 1 and Table 4) when they are both averaged in spatial bins and compared as a function of hypocentral distance. The correlation improves with the increasing number of MyShake felt reports, we find 10 reports in distance bins can be aggregated to a stable measurement. The spatially averaged felt reports can also be plotted as a map to provide a shaking intensity map. Through this established link between MyShake felt reports and MMI, the crowdsourced MyShake felt reports can be used by the scientific community to provide another independent shaking intensity dataset.

**Data and Resources**

MyShake data are currently archived at Berkeley Seismology Laboratory and use is constrained by the privacy policy of MyShake (see http://myshake.berkeley.edu/privacy-policy/index.html), but data for the research purposes can be requested from the authors. The data for the Gridded Population of the World can be accessed at https://sedac.ciesin.columbia.edu/data/collection/gpw-v4. USGS DYFI data can be accessed at https://earthquake.usgs.gov/data/dyfi/. USGS ANSS Comprehensive Earthquake Catalog (ComCat) can be accessed at https://earthquake.usgs.gov/earthquakes/search/.

**Acknowledgements**

The Gordon and Betty Moore Foundation funded this analysis through grant GBMF5230 to UC Berkeley. The California Governor's Office of Emergency Services (Cal OES) funds the operation of MyShake through grant 6142-2018 to Berkeley Seismology Lab. We thank the two anonymous reviewers and the editor Allison Bent for their constructive comments and suggestions, which improves the paper quality. We thank Vince Quitoriano and David Wald from USGS for sharing code utilities for the UTM boxes. We thank the previous and current MyShake team members: Roman Baumgaertner, Garner Lee, Arno Puder, Louis Schreier, Stephen Allen, Stephen Thompson, Jennifer Strauss, Kaylin Rochford, Akie Mejia, Doug Neuhauser, Stephane Zuzlewski, Asaf Inbal, Sarina Patel and Jennifer Taggart. All the analysis of this project is done in Python. We thank all the MyShake user citizen scientists for their data contributions. We also thank the USGS DYFI project for enabling this study. Qingkai Kong's work was performed under the auspices of the U.S. Department of Energy by Lawrence Livermore National Laboratory under

**Tables, with captions above each table**

Table 1 Questions used in MyShake felt report questionnaires

| Question Number | Question | User can choose from |
|---|---|---|
| 1 | Where were you when you experienced this earthquake? | Click on map or type in address |
| 2 | What strength of shaking did you feel? | None, light, moderate, strong, severe |
| 3 | Describe any visible building damage that you see nearby? | No, minor, substantial, destroyed |
| 4 | Describe any visible road damage that you see nearby? | No, minor, substantial, destroyed |

Table 2. An example of the felt report data used in the study.

|  | **Type** | **Example** |
|---|---|---|
| **Time of the submission** | Integer, Unix timestamp | 1621007178230 |

| Location of the report | Float Latitude and Longitude pair | (37.23, -122.34) |
|---|---|---|
| **Shaking level** | Integer: -1, 0, 1, 2, 3 | 2 |

Table 3. List of events that have more than 700 felt reports submitted, the bolded events are reserved for testing purposes.

| **Earthquake id** | **Date and Time** | Magnitude | # of reports |
|---|---|---|---|
| ci38695658 | 2020-09-19 06:38:46 | 4.5 | 9829 |
| **nc73512355** | **2021-01-17 04:01:27** | **4.2** | **4810** |
| nc73291880 | 2019-10-15 05:33:42 | 4.5 | 4432 |
| **ci39838928** | **2021-04-05 11:44:01** | **4** | **4357** |
| ci39400304 | 2020-04-22 07:03:47 | 3.7 | 3723 |
| ci39126079 | 2020-04-04 01:53:18 | 4.9 | 3050 |
| ci39462536 | 2020-06-04 01:32:11 | 5.5 | 2789 |
| ci39277736 | 2020-01-22 07:41:10 | 3.6 | 2787 |
| ci39493944 | 2020-06-24 17:40:49 | 5.8 | 2494 |
| ci38905415 | 2019-10-18 07:19:51 | 3.5 | 2211 |
| ci39322287 | 2020-07-30 11:29:29 | 4.2 | 1808 |
| nc73559265 | 2021-05-07 04:35:14 | 4.7 | 1671 |
| nc73322626 | 2020-01-02 07:16:31 | 3.9 | 1354 |
| nc73505175 | 2020-12-31 13:41:59 | 3.3 | 1207 |
| nc73510910 | 2021-01-14 19:18:10 | 3.6 | 1187 |
| nn00725272 | 2020-05-15 11:03:27 | 6.5 | 931 |
| **ci39322767** | **2020-07-30 13:48:19** | **3.7** | **836** |
| **nc73292360** | **2019-10-15 19:42:30** | **4.7** | **760** |
| nc73554215 | 2021-04-25 04:59:28 | 3.6 | 748 |
| ci39762912 | 2021-01-20 16:31:58 | 3.5 | 705 |

Table 4 Regression results, t shows the t statistics, and P > |t| is the corresponding p-value. The [0.025 0.975] interval shows the 95% confidence interval.

| | coef | std err | t | P > |t| | [0.025 | 0.975] |
|---|---|---|---|---|---|---|
| a | 2.30 | 0.044 | 51.85 | 0.00 | 2.21 | 2.39 |
| b | 2.44 | 0.129 | 18.97 | 0.00 | 2.19 | 2.69 |

**List of figure captions**

**Figures, with captions below each figure**

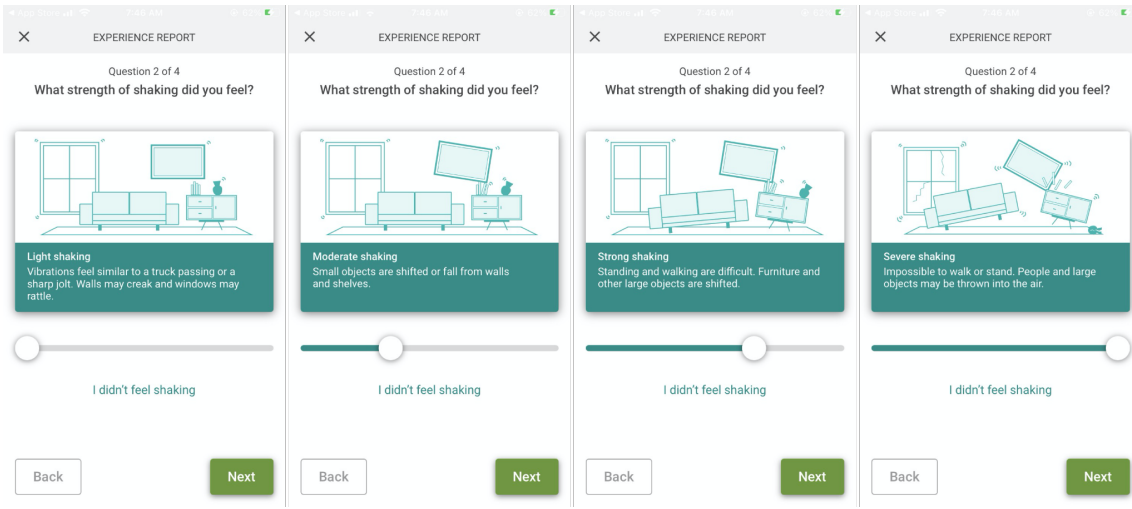

Figure 1. The 5 different shaking levels that users can select within MyShake felt report questionnaire (none, light, moderate, strong and severe). Users scroll through the images shown to identify the one that best matches his/her experience.

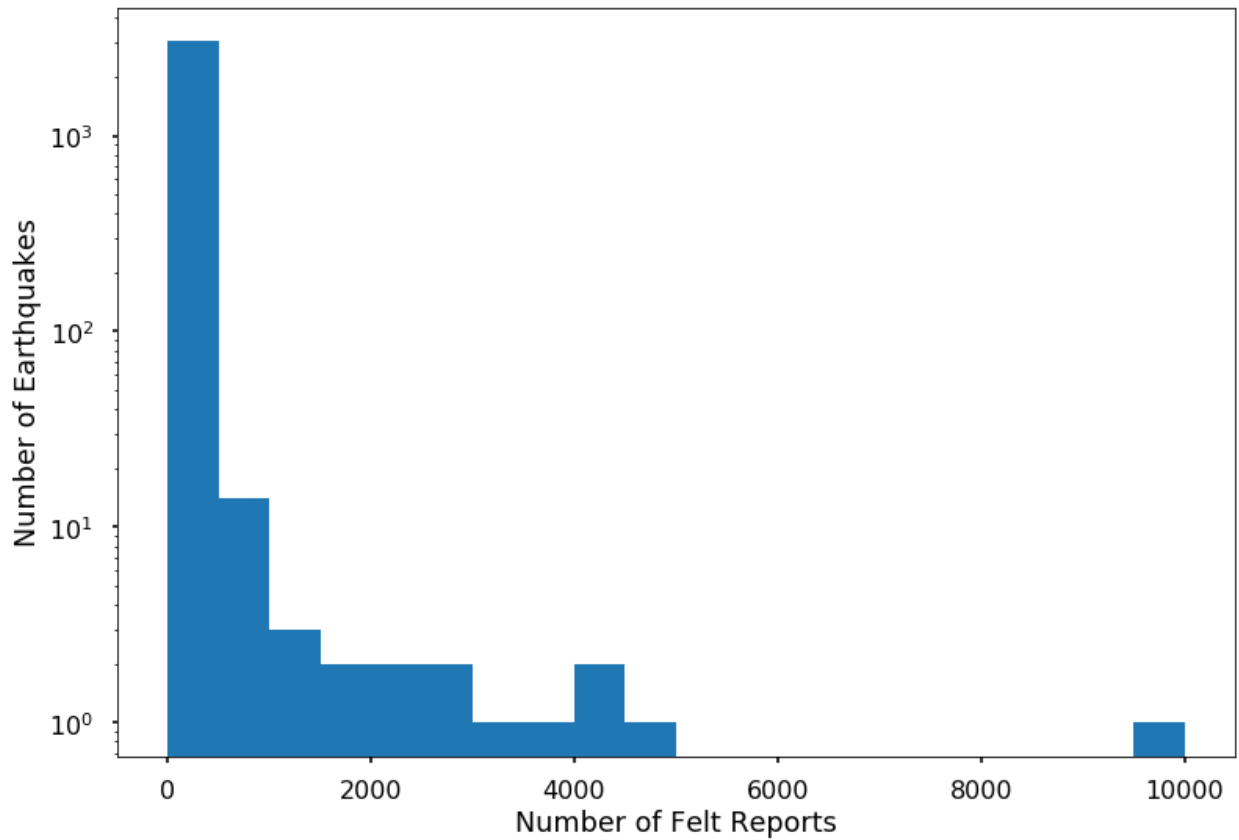

Figure 2. Histogram showing the number of earthquakes for which a specified number of felt reports were submitted between 2019-10-15 and 2021-05-11 in California.

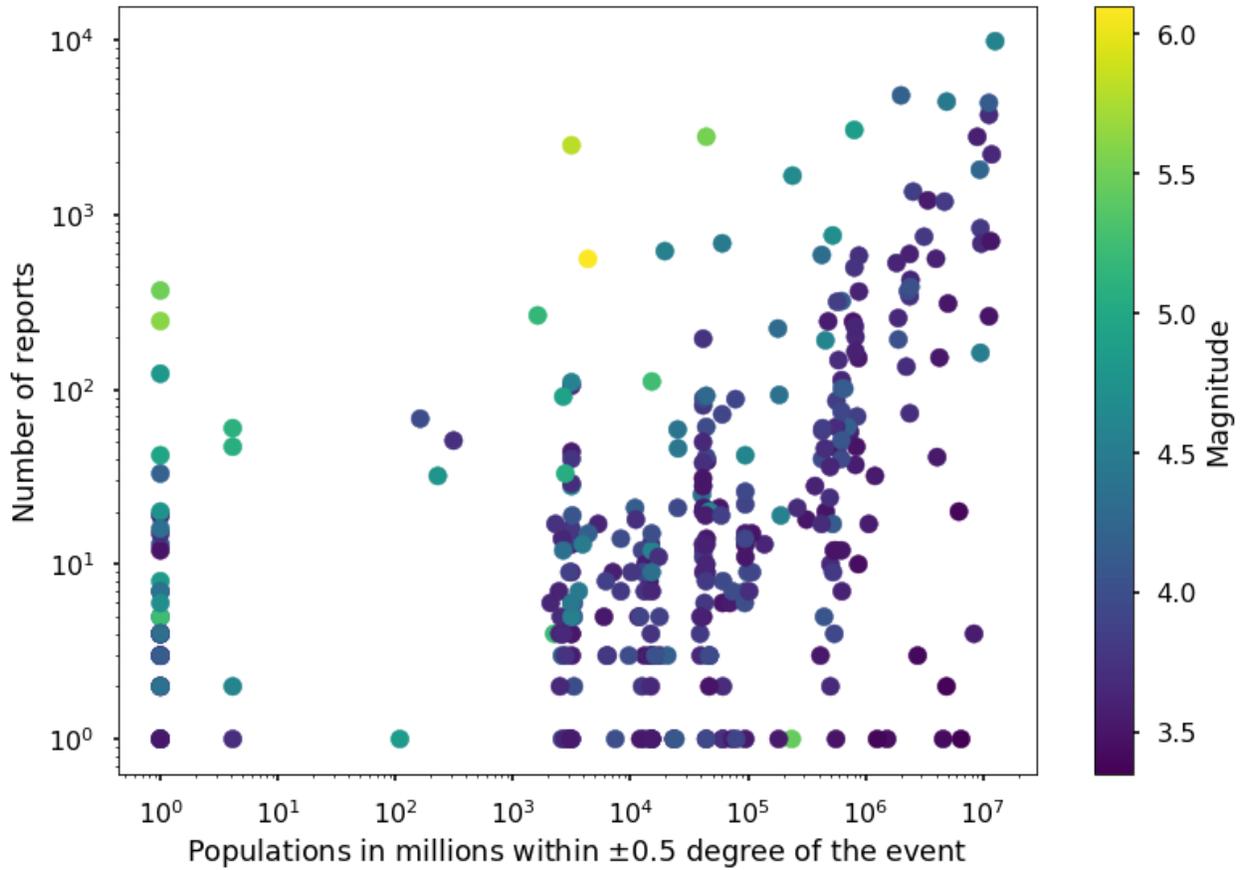

Figure 3. Number of felt reports versus population (in millions) with 0.5 degree of the earthquake. The color shows the magnitude of the earthquake. Note, we add 1 to the reported population for each earthquake in order to plot on a logarithmic scale.

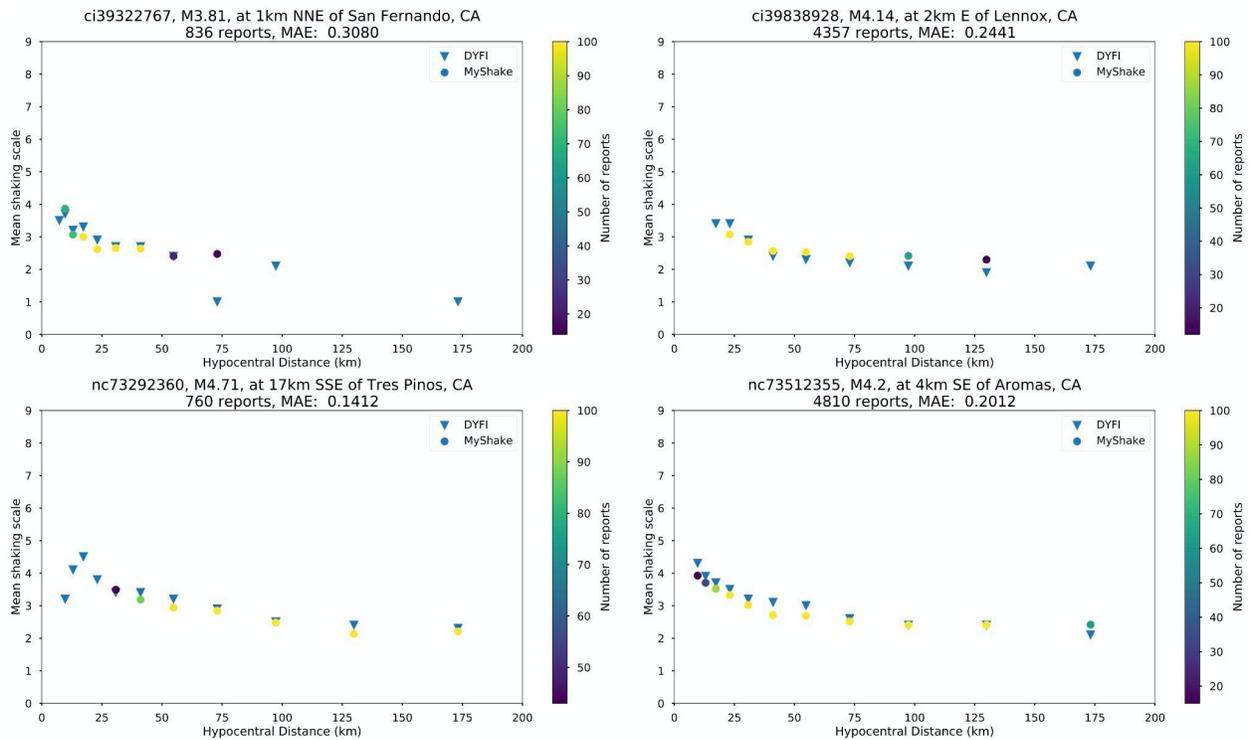

Figure 4. Shaking intensity (MMI scale) versus distance for four earthquakes in the test set. Each panel compares the estimated MMI shaking derived from the MyShake felt reports using equation 1 to the MMI estimates from DYFI. The MyShake data are shown as circles, and the DYFI data are shown as inverted triangles. The color of the circles represents the number of reports in each distance bins. The distance bins are set to be the same as reported by DYFI: 13.0, 17.3, 23.1, 30.8, 41.1, 54.8, 73., 97.4, 129.9, 173.2, and 200 km. Each figure title gives the event id, magnitude, place of the earthquake, number of felt reports and Mean Absolute Error (MAE). The bins with fewer than 10 MyShake felt reports are not plotted.

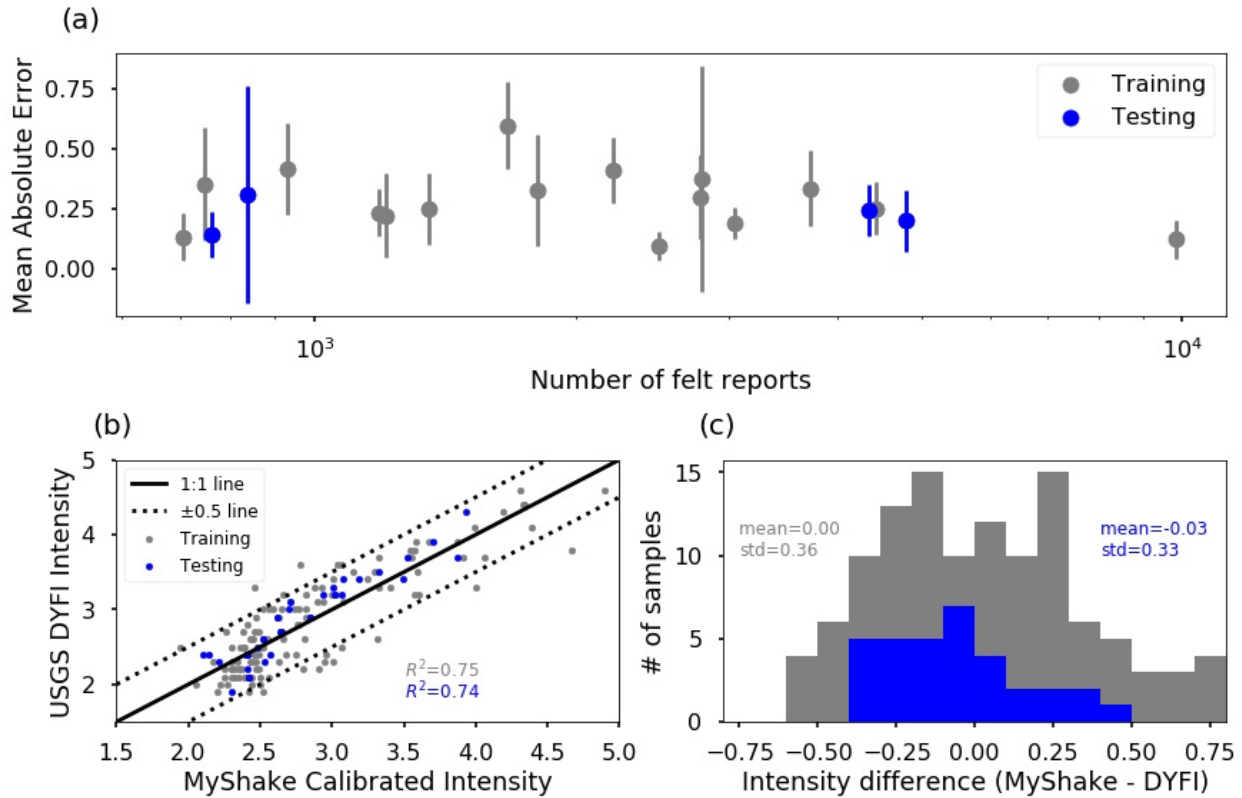

Figure 5. Fit metrics between MyShake calibrated intensity and the DYFI intensity in different distance bins for the 20 events with the most felt reports. (a) Mean Absolute Errors versus the number of felt reports for each event. The vertical bars are the standard deviation within each event. (b) Scatter plot between USGS DYFI and MyShake calibrated intensities. The solid line and the two dotted lines are 1-to-1 and the 0.5-unit error lines, respectively. The R-squared value is the coefficient of determination. (c) The histogram of the intensity difference between the MyShake calibrated and DYFI, the mean and standard deviation are listed in the figure.

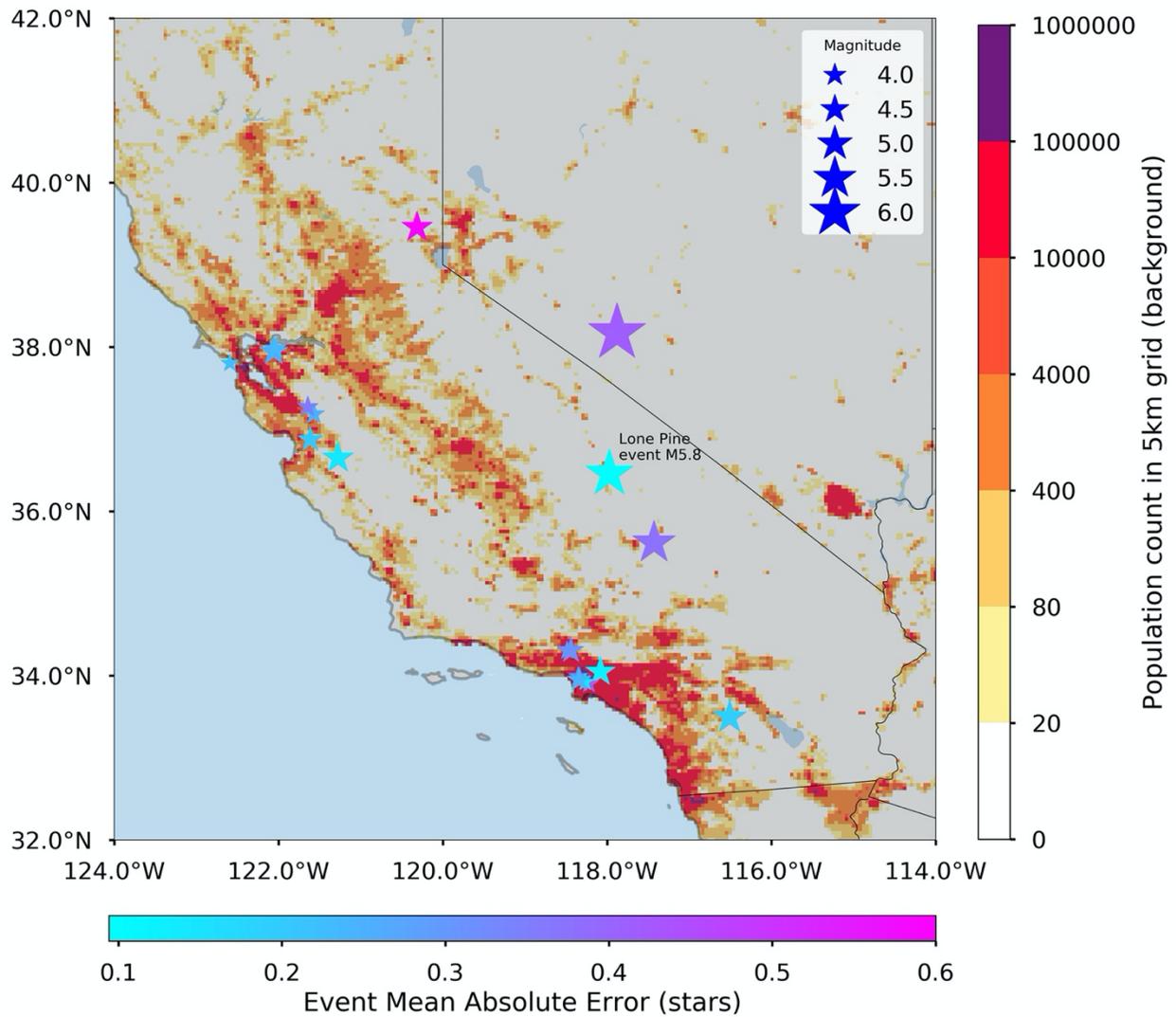

Figure 6. Spatial distribution of the 20 events with more than 700 felt reports. The size of the star represents magnitude, and the color represents the MAE. The population counts in 5 km grid are plotted on the map background.

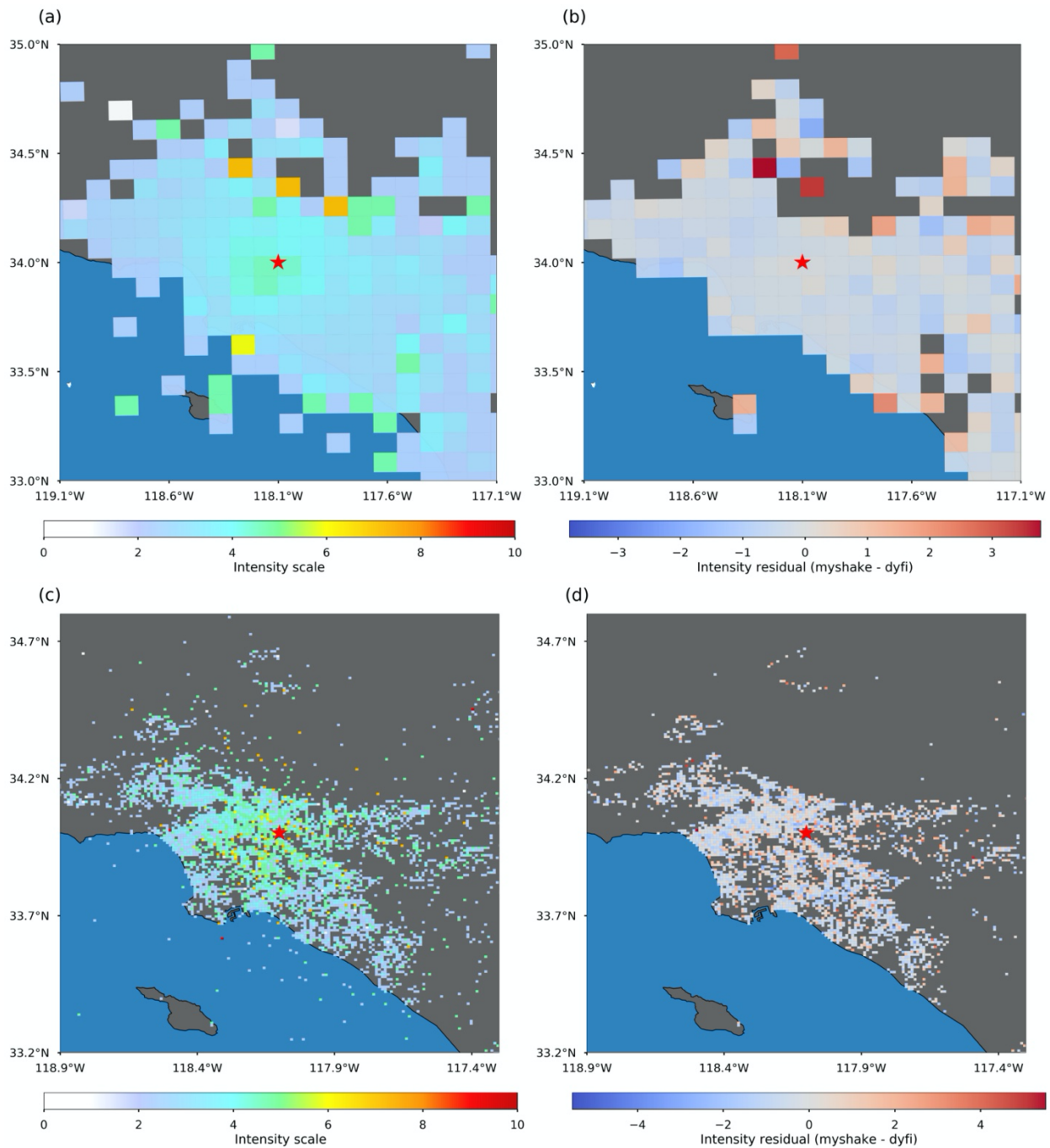

Figure 7. Spatial Intensity distribution and the residual compared to DYFI in 10 km and 1 km UTM boxes. (a) Derived calibrated intensity map from MyShake felt reports in 10 km UTM boxes. The map uses the same intensity color scale as the one used in ShakeMap by the USGS. (b) The intensity residual map compared to 10 km DYFI from USGS, the mean and standard deviation of the residuals are -0.08 and 0.75, respectively. (c, d) Same as (a) and (b), but for 1 km UTM boxes. The mean and standard deviation of the residuals for the 1 km UTM boxes are -0.22 and 1.12, respectively. Note, residual maps are made only for places where data available for both MyShake and DYFI. Due to the small number of MyShake felt reports in each UTM box, we don't use any quality control for this map, i.e. no requirement of minimum number of reports in each box to average. The red star is the location of the 2020-09-19 M4.5 earthquake. The corresponding DYFI intensity map is shown in figure S12.

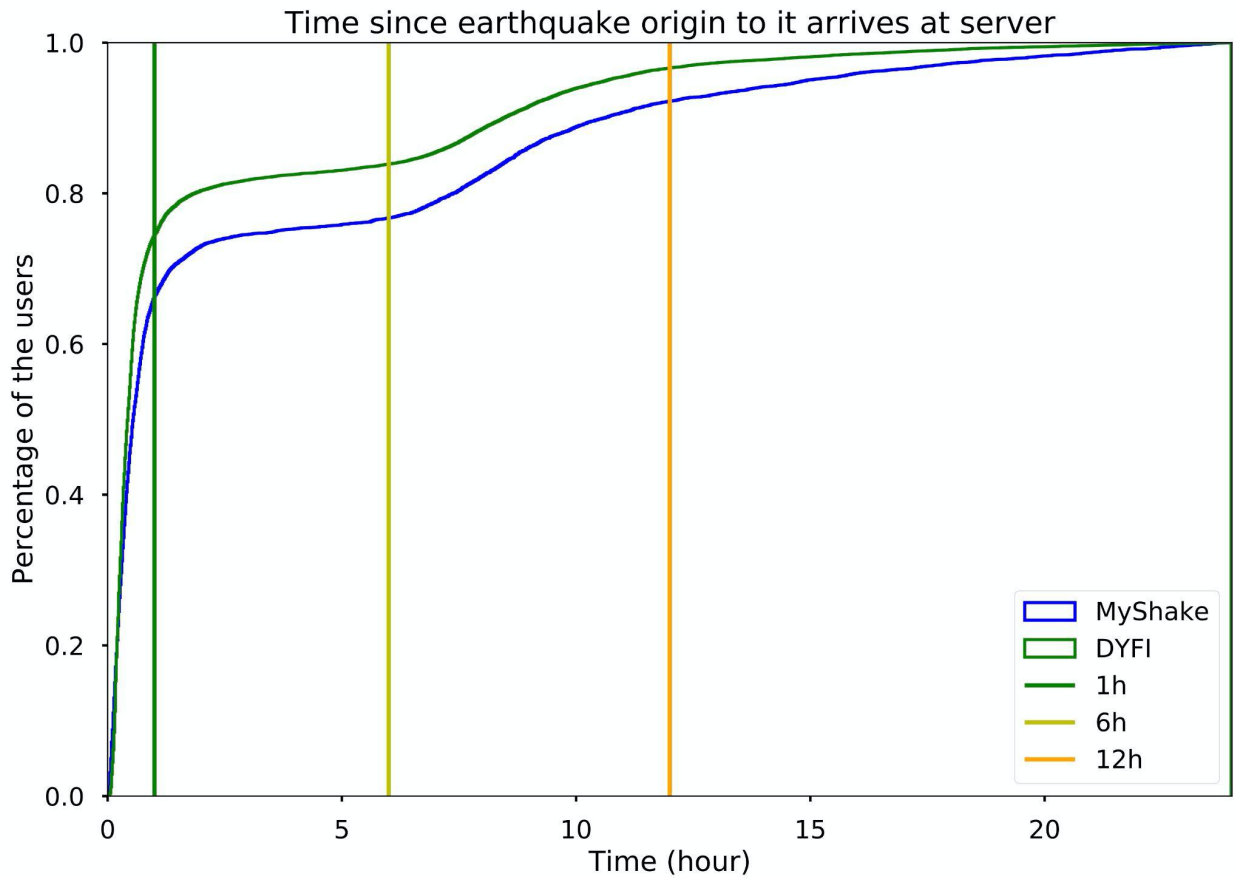

Figure 8. Reporting time history for the MyShake and DYFI felt reports. The percentage of the felt reports is shown versus time after the origin of the earthquake. The 1-hour, 6-hour and 12-hour lines are also plotted in the figure.

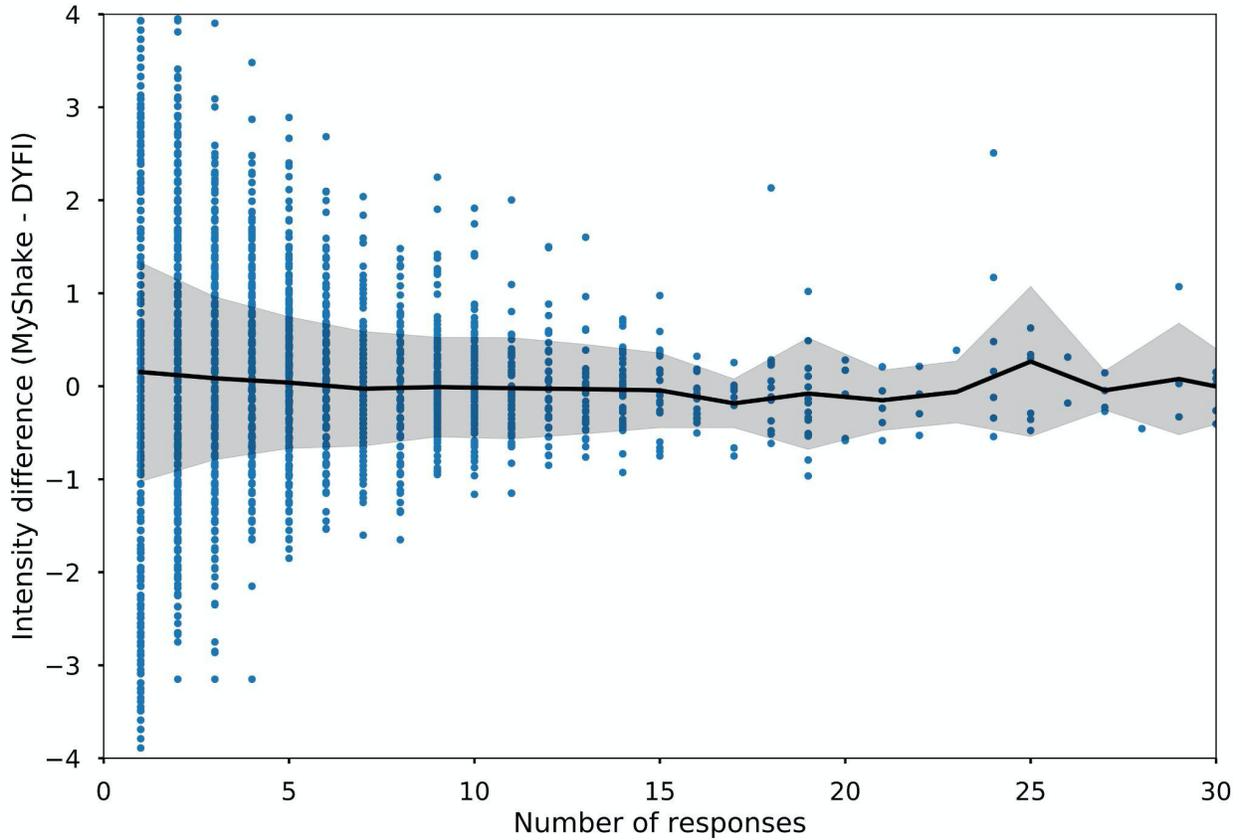

Figure 9. Number of responses versus intensity difference between calibrated MyShake and USGS DYFI. The blue dots are the intensity differences within the 1km UTM boxes, the solid black line and the gray shaded area are the mean and standard deviation of the intensity difference.

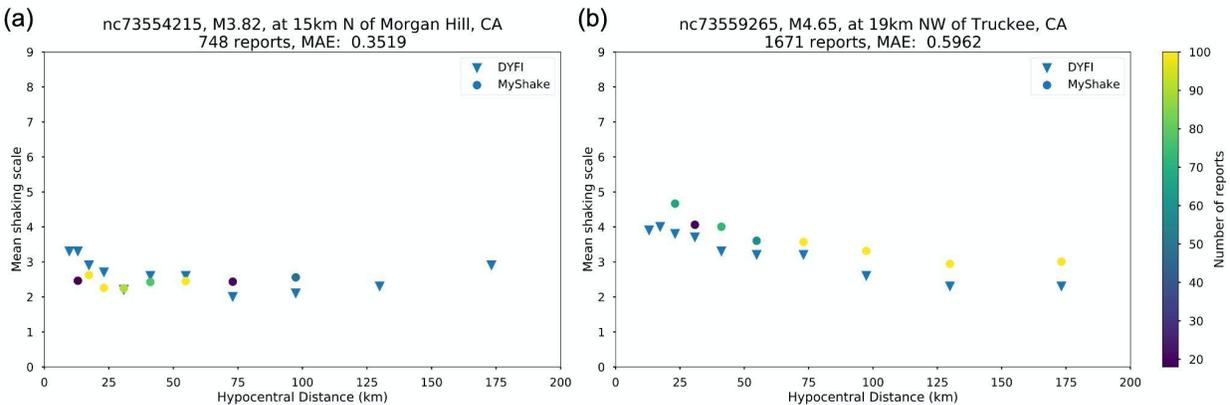

Figure 10. Two events with large intensity errors between the MyShake and DYFI felt reports. (a) M3.86 Morgan Hill event (b) M4.65 Truckee event. MyShake shaking scales are color coded by the number of felt reports.

# Supplementary Material for Crowdsourcing Felt Reports using the MyShake smartphone app

Qingkai Kong, Richard M Allen, Steve Allen, Theron Bair, Akie Meja, Sarina Patel, Jennifer Strauss, Stephen Thompson.

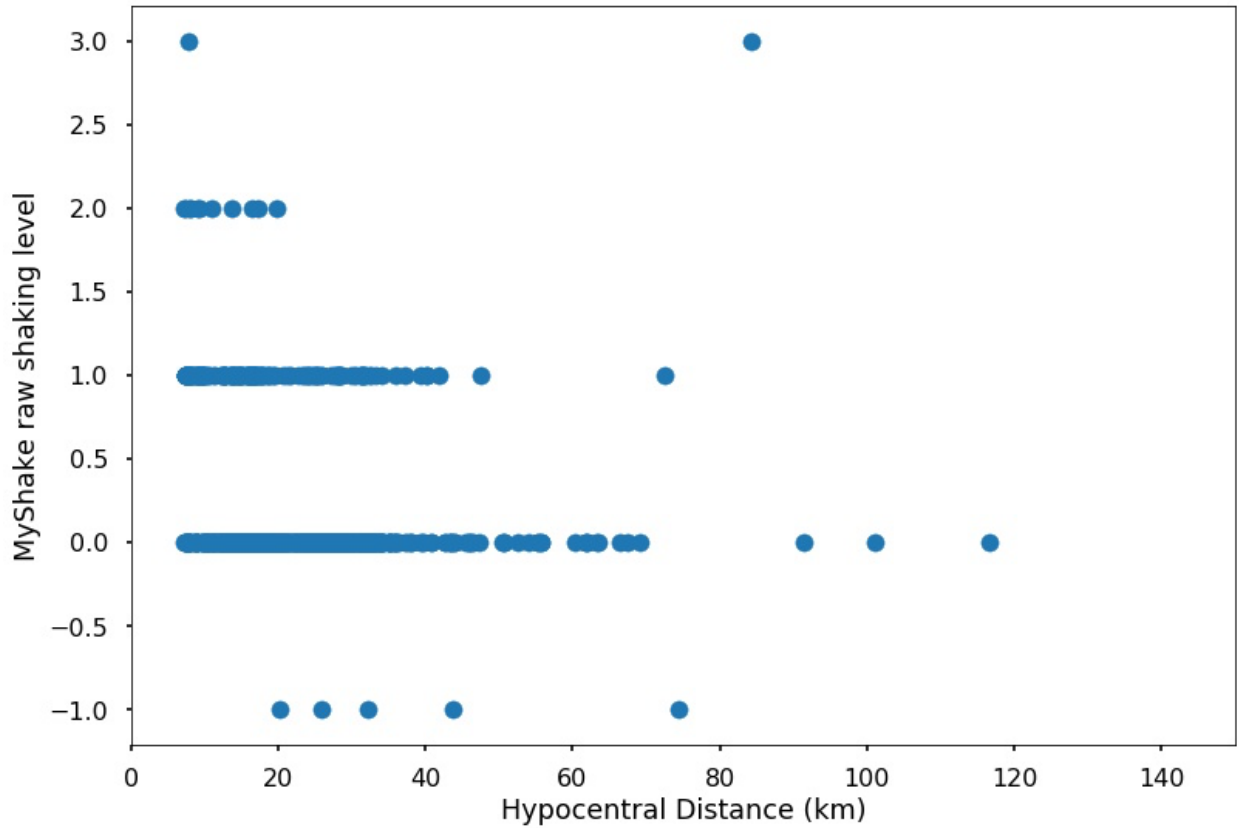

Figure S1 Example of the raw felt reports from MyShake for the 2020-07-30 M3.7 San Fernando earthquake (id: ci39322767). Users report the shaking intensity on the 5-point scale and the information is recorded along with their location.

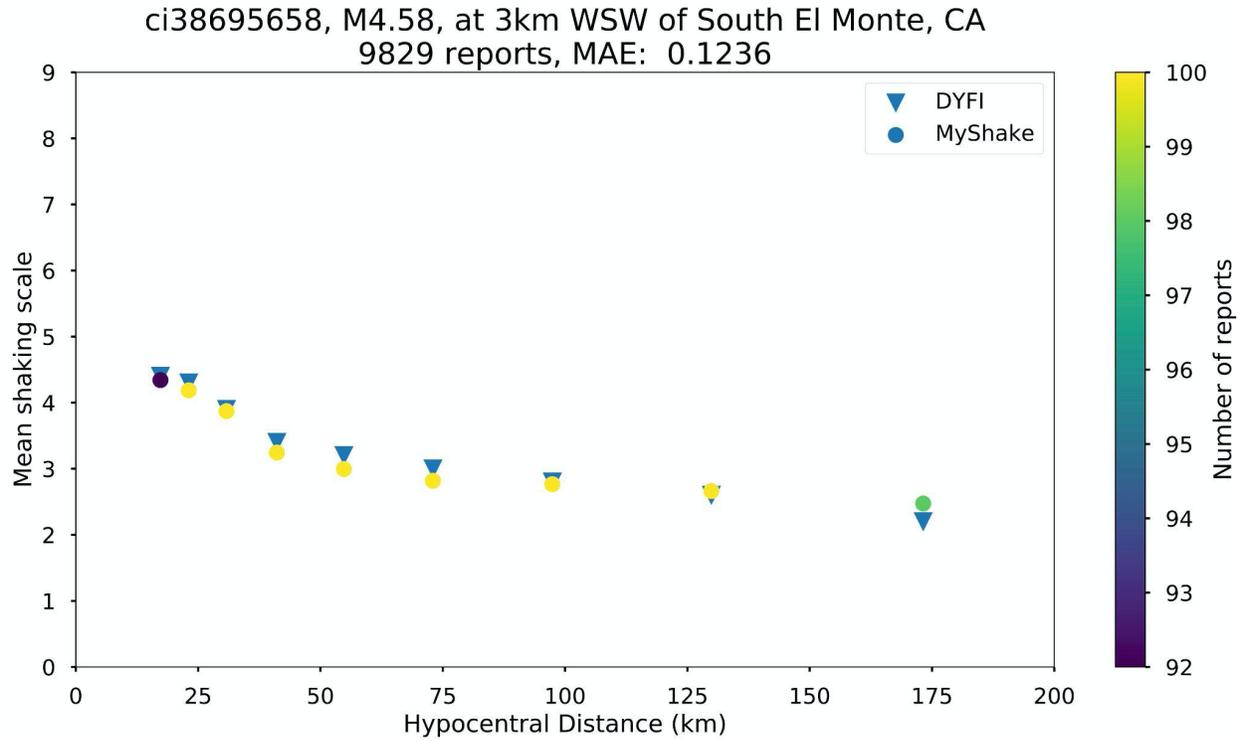

Figure S2 - Shaking scale comparison between MyShake converted MMI with DYFI MMI for event ci38695658

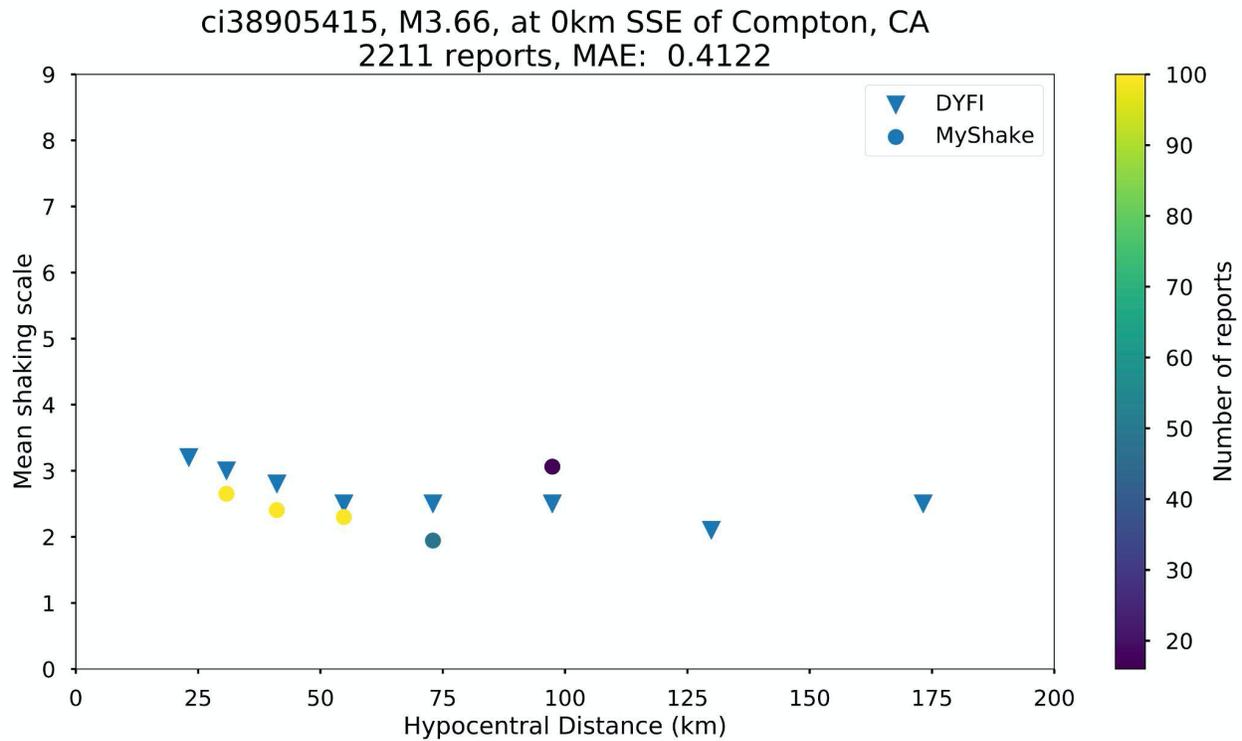

Figure S3 - Shaking scale comparison between MyShake converted MMI with DYFI MMI for event ci38905415

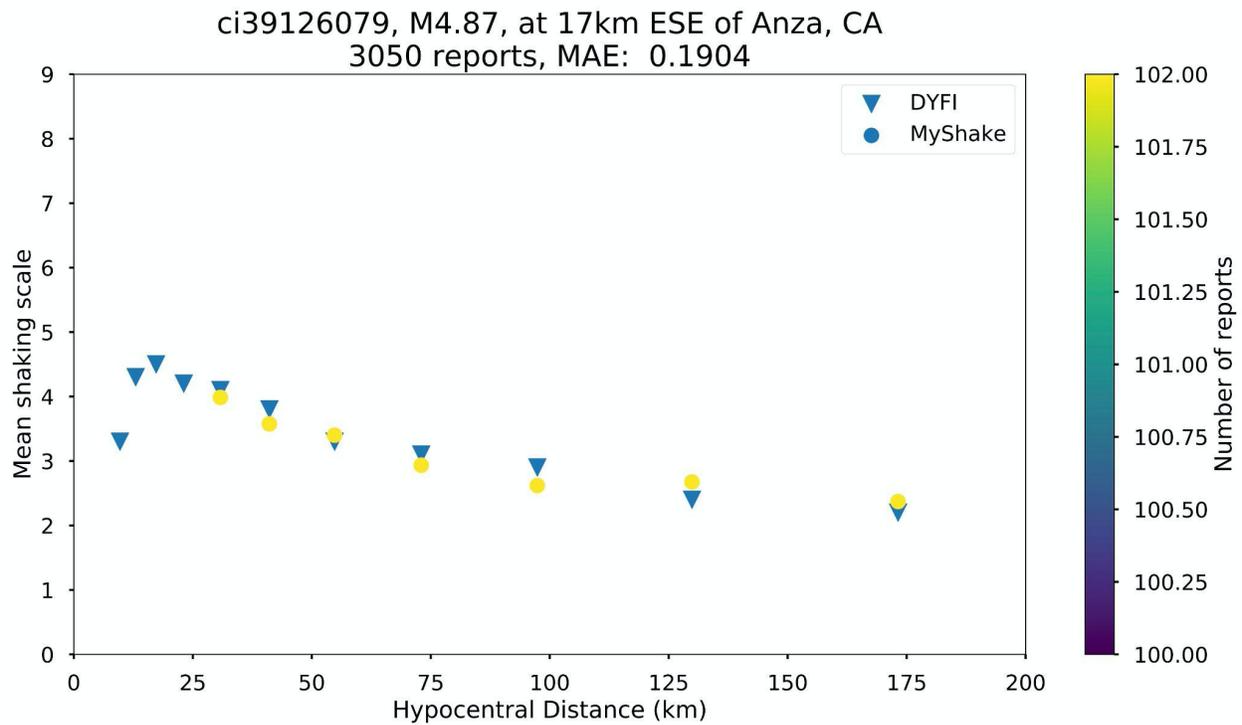

Figure S4 - Shaking scale comparison between MyShake converted MMI with DYFI MMI for event ci39126079

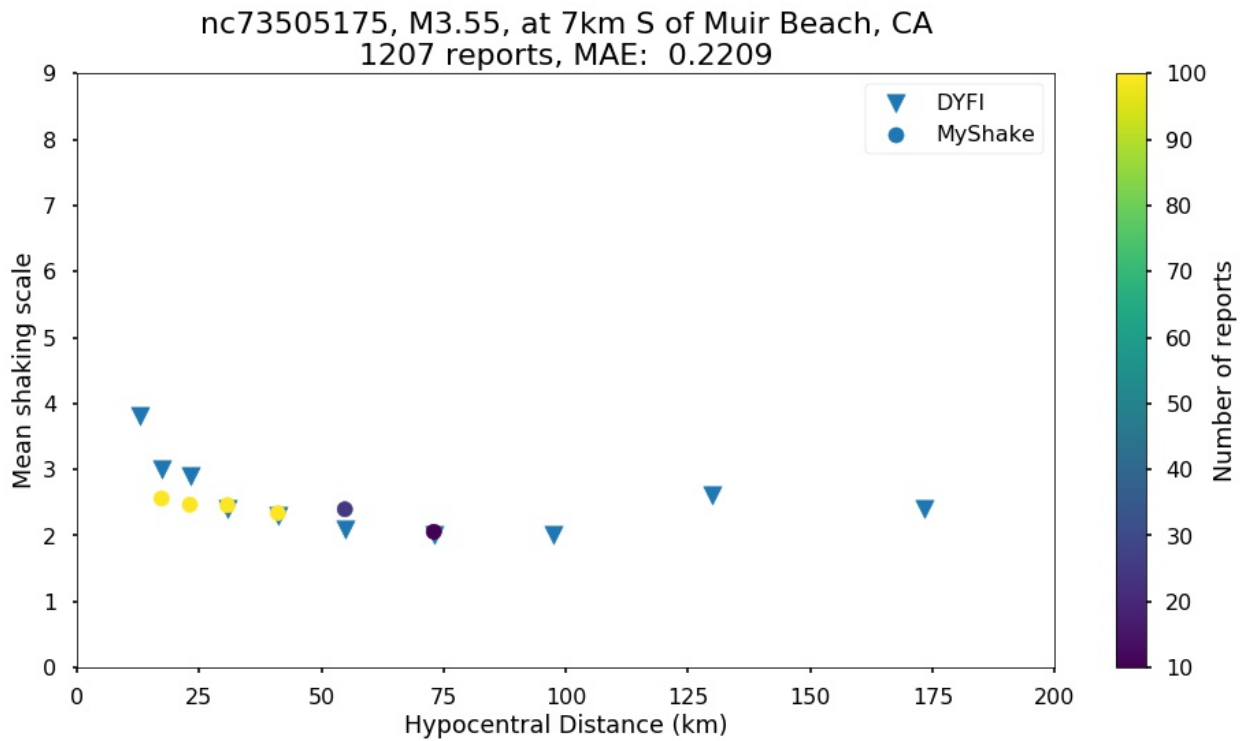

Figure S5 - Shaking scale comparison between MyShake converted MMI with DYFI MMI for event nc73505175

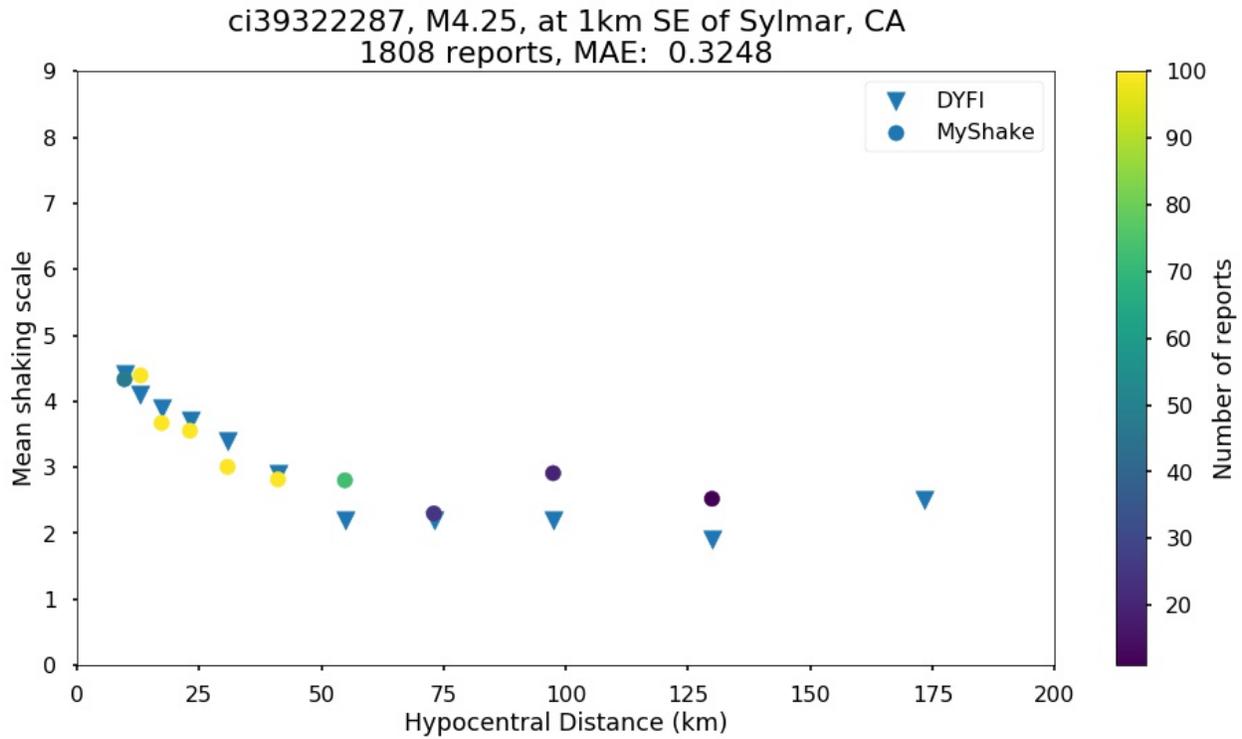

Figure S6 - Shaking scale comparison between MyShake converted MMI with DYFI MMI for event ci39322287

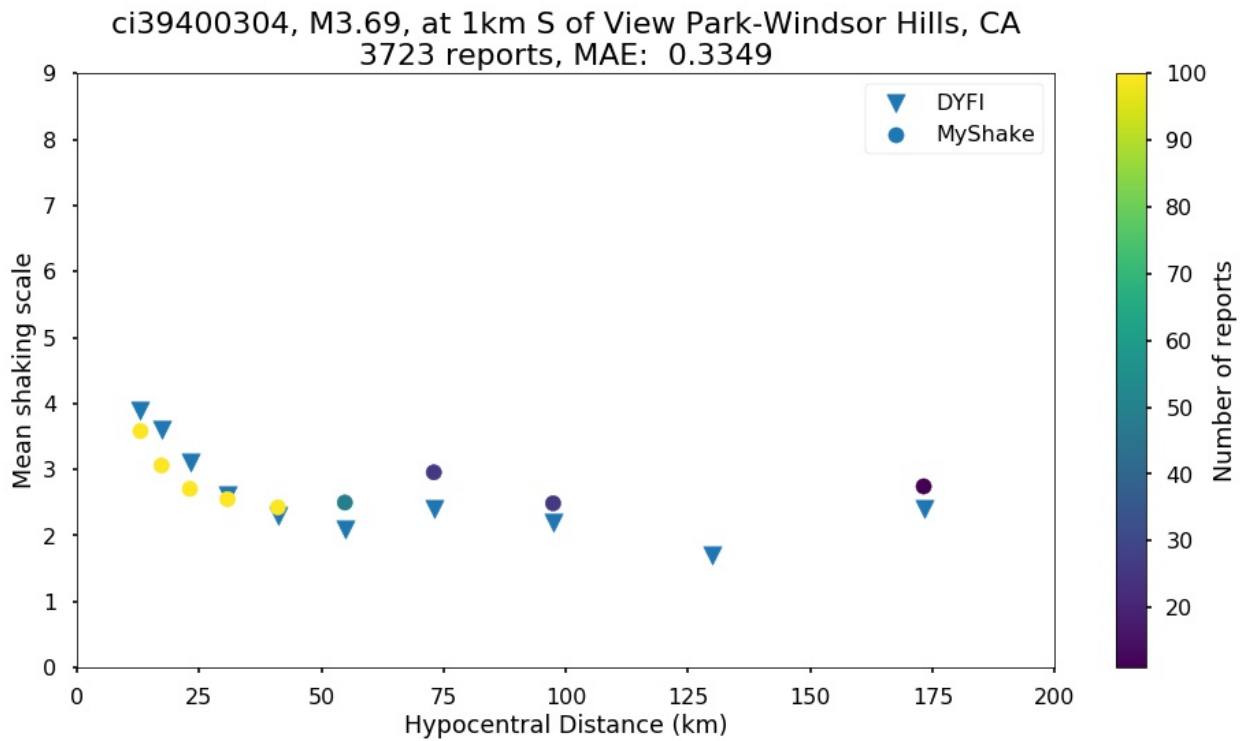

Figure S7 - Shaking scale comparison between MyShake converted MMI with DYFI MMI for event ci39400304

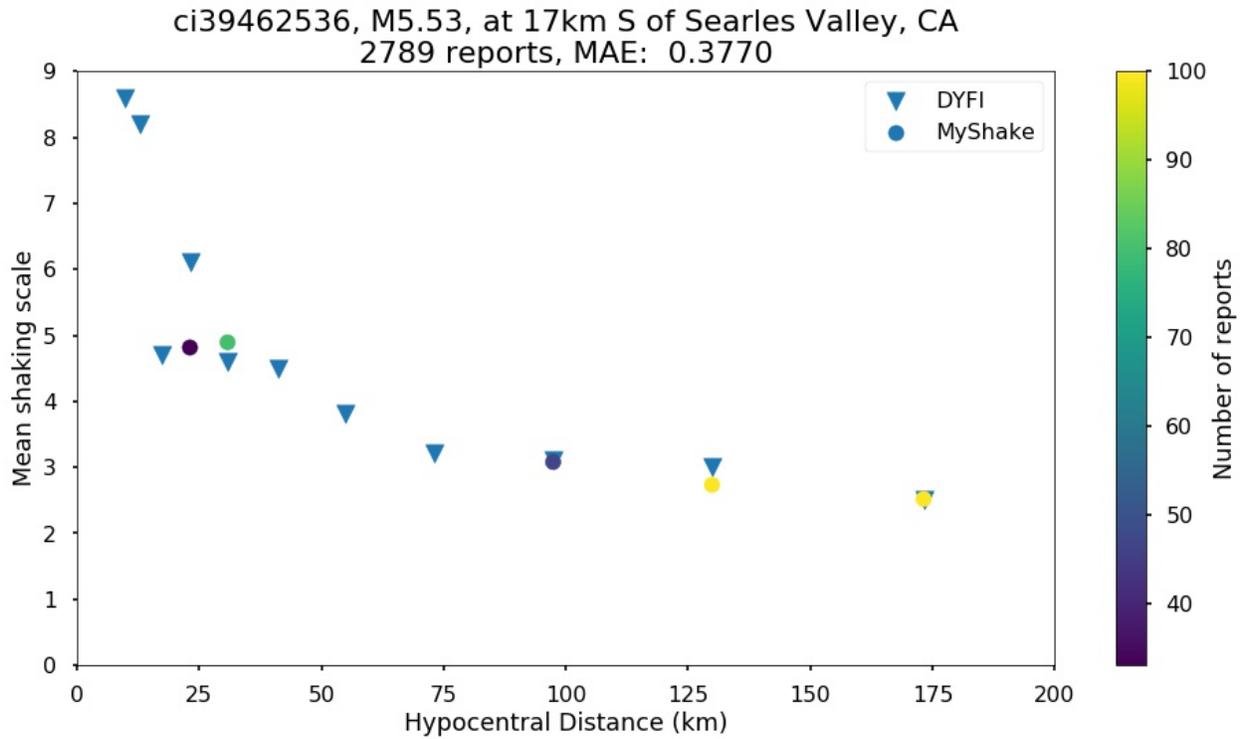

Figure S8 - Shaking scale comparison between MyShake converted MMI with DYFI MMI for event ci39462536

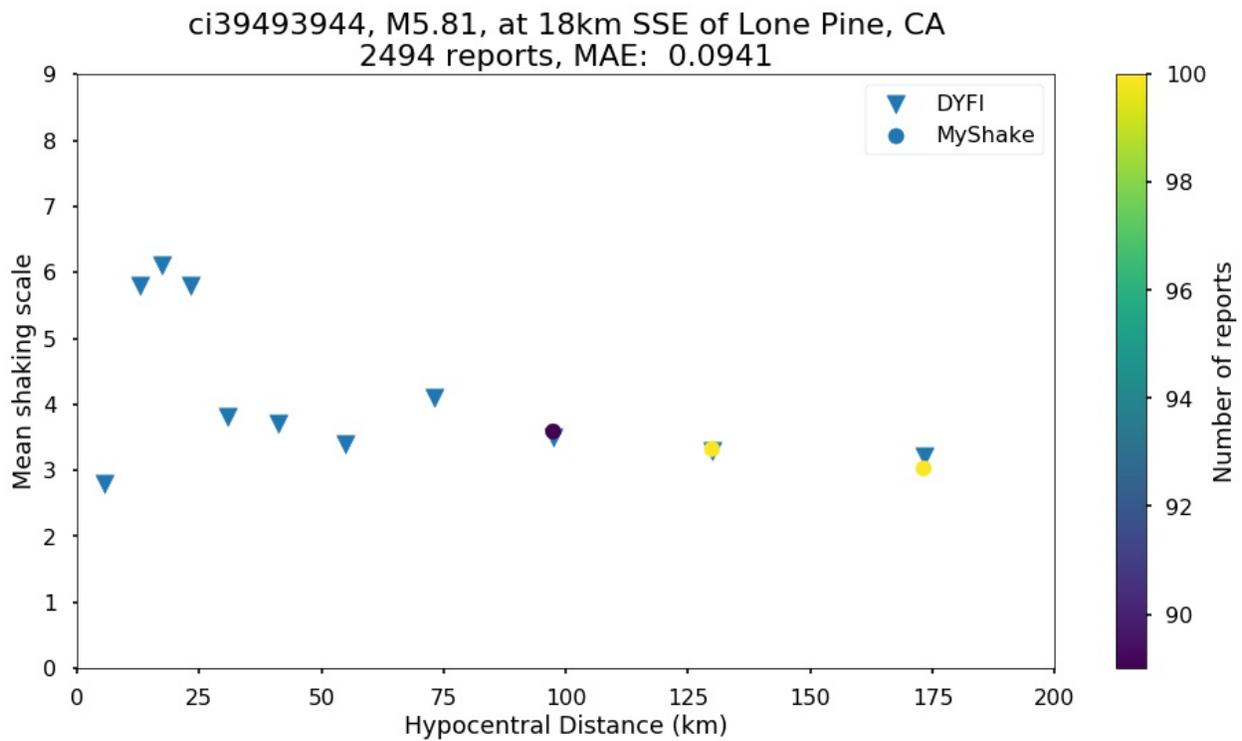

Figure S9 - Shaking scale comparison between MyShake converted MMI with DYFI MMI for event ci39493944

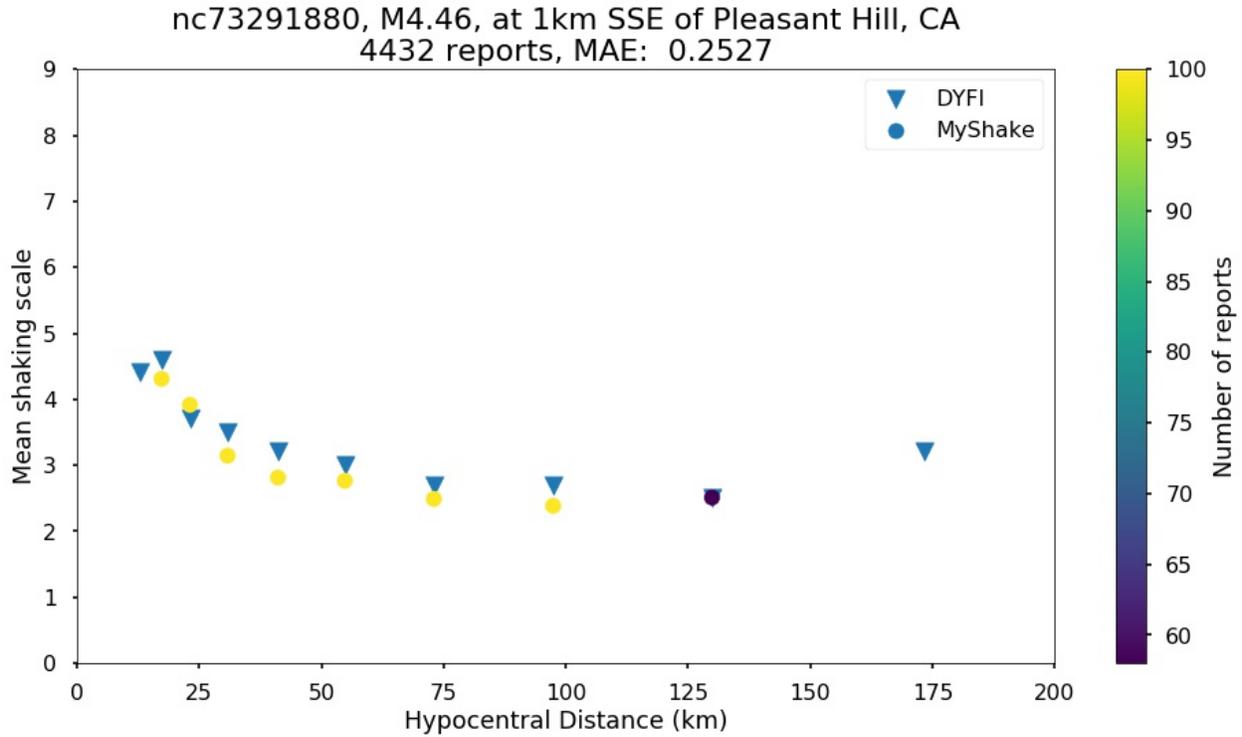

Figure S10 - Shaking scale comparison between MyShake converted MMI with DYFI MMI for event nc73291880

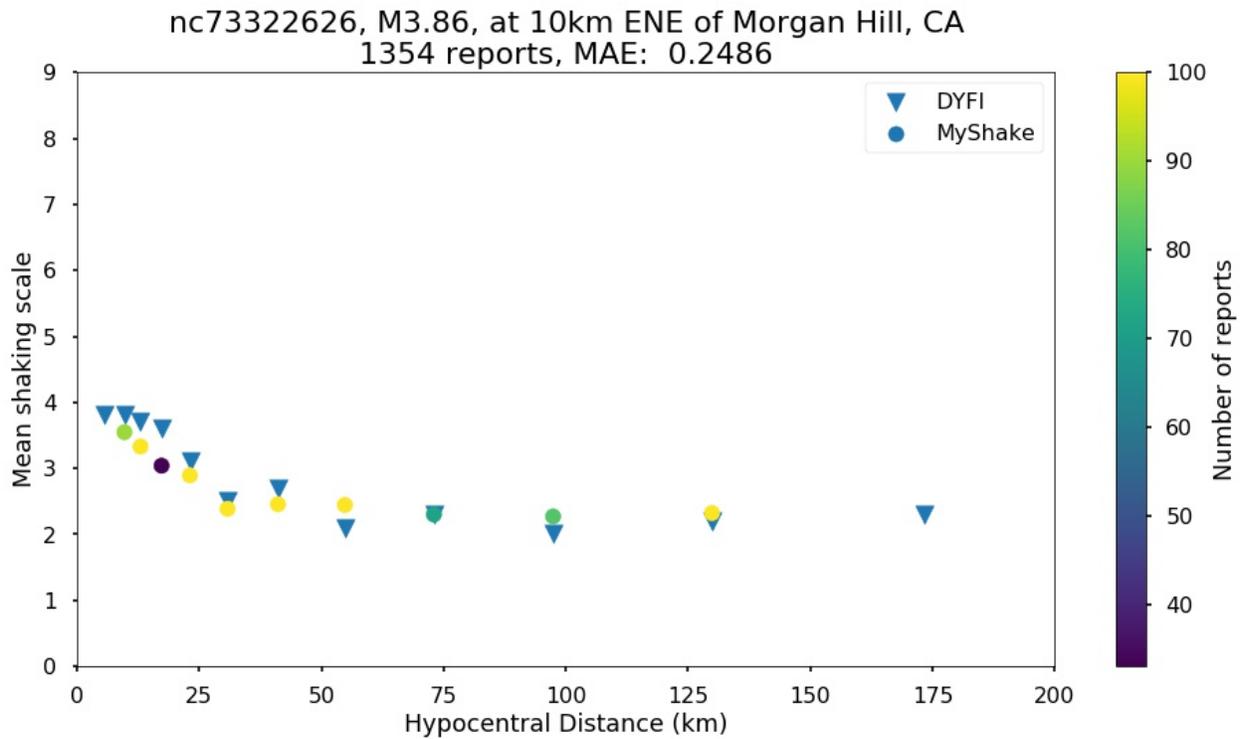

Figure S11 - Shaking scale comparison between MyShake converted MMI with DYFI MMI for event nc73322626

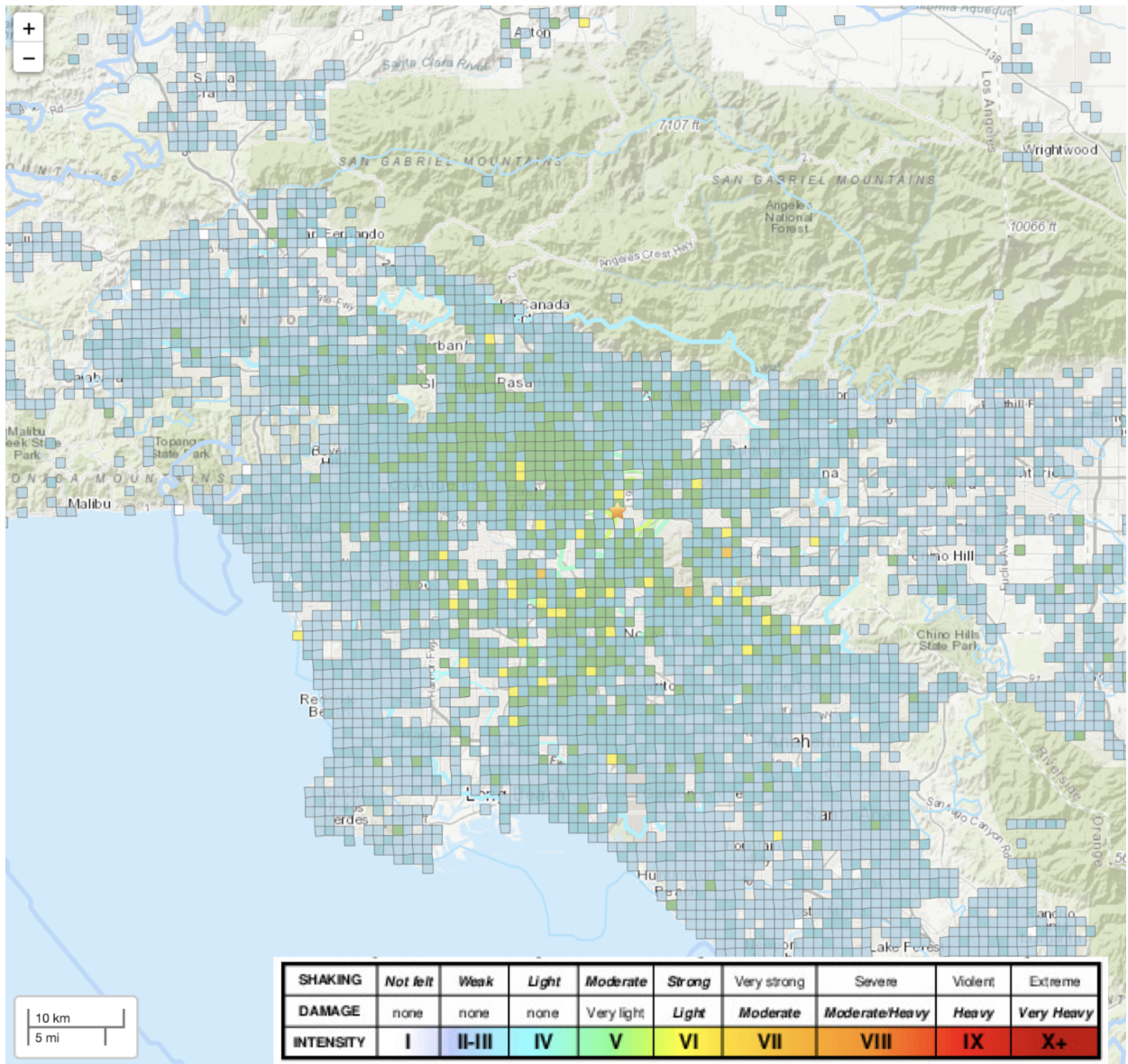

Figure S12 - The 1km DYFI intensity map for 2020-09-19 M4.5 earthquake. The corresponding intensity map using MyShake felt reports is shown in figure 5 in the main text.

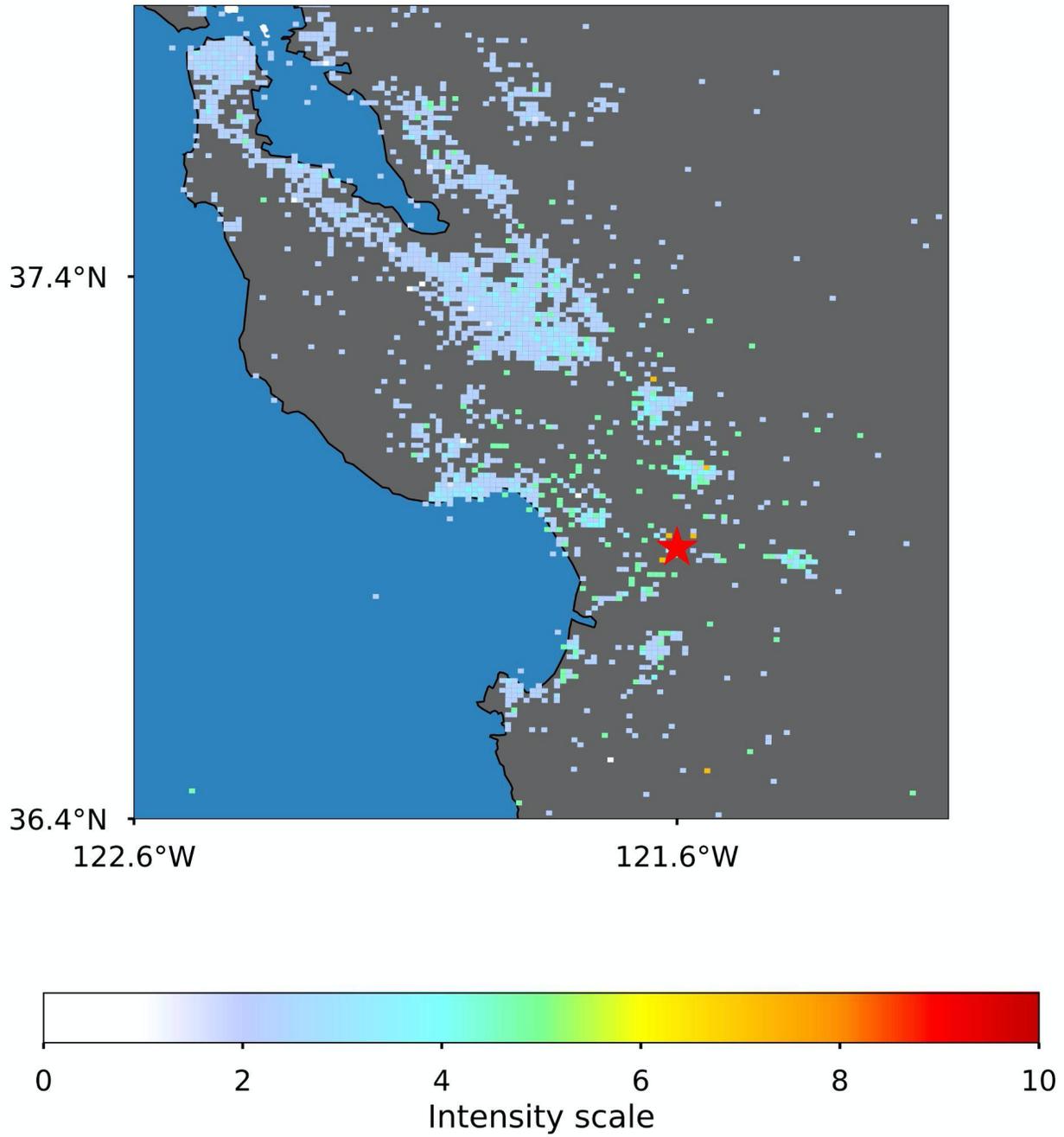

Figure S13 - Derived intensity map from MyShake converted MMI scale for 2021-01-17 M4.2 earthquake (nc73512355), we use the same color scale and spatial boxes (1km UTM boxes) as the one used in ShakeMap from USGS.

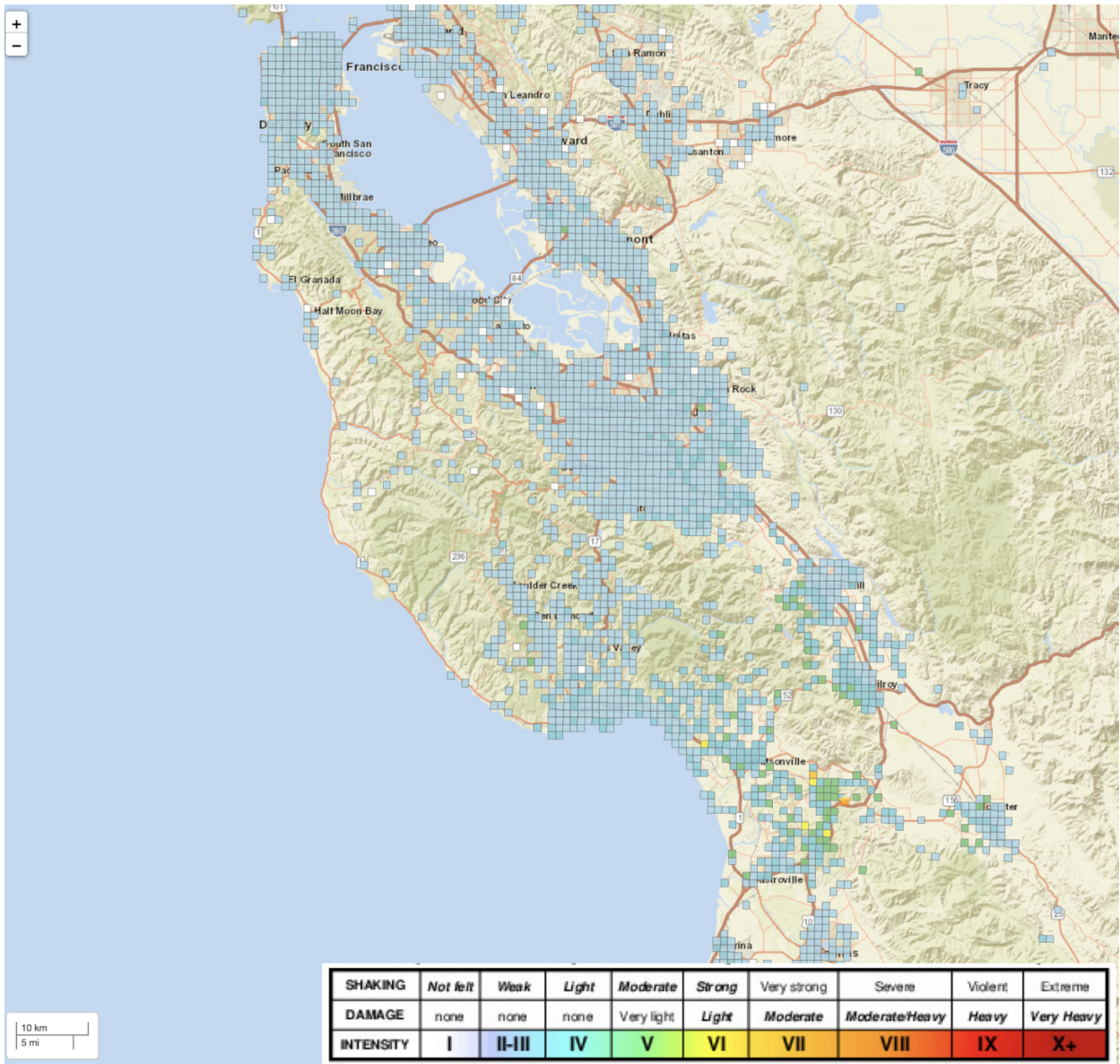

Figure S14 - Corresponding MMI scale for 2021-01-17 M4.2 earthquake (nc73512355) from USGS DYFI.

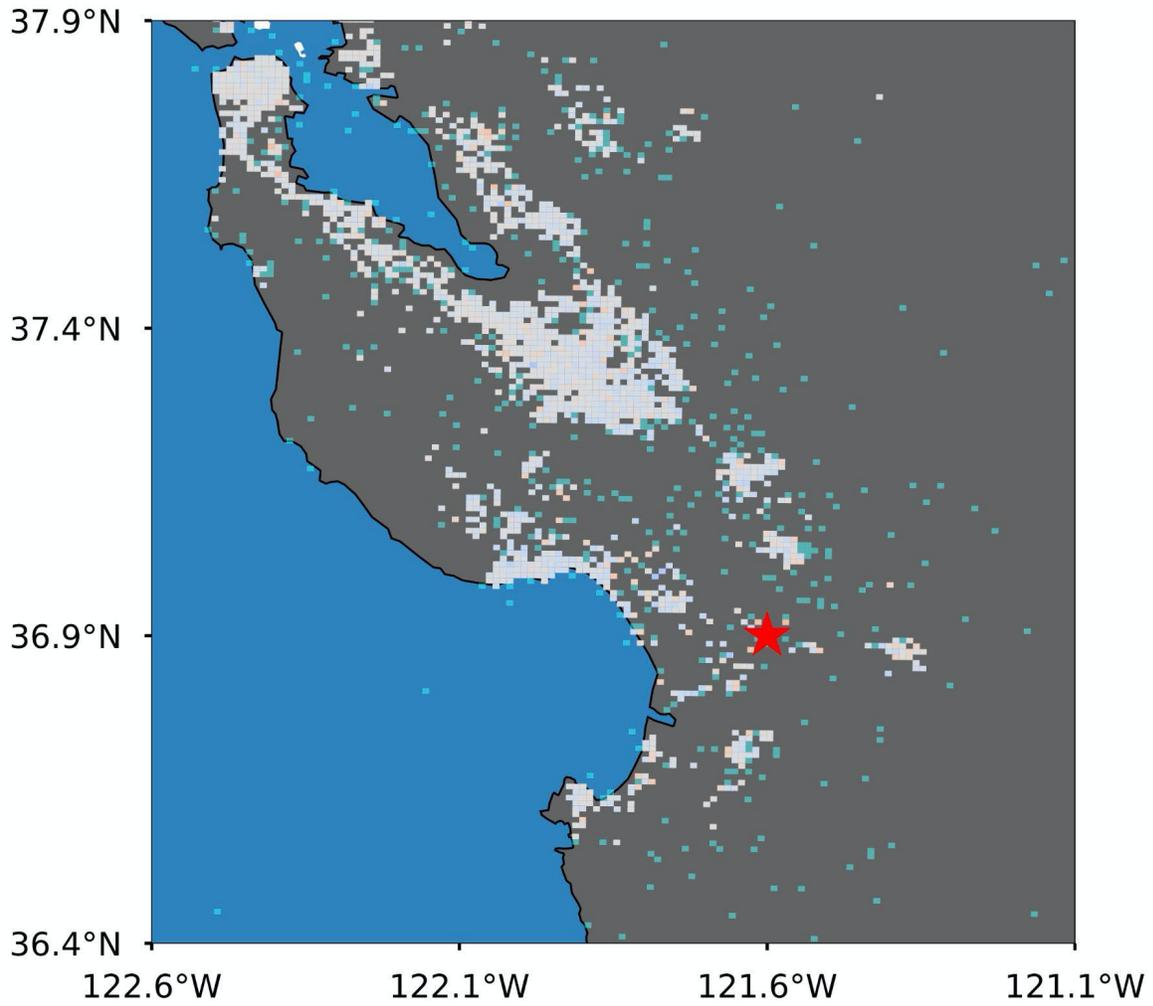

Figure S15 - Residual (MyShake - DYFI) in the 1km UTM boxes for 2021-01-17 M4.2 earthquake (nc73512355) from USGS DYFI.

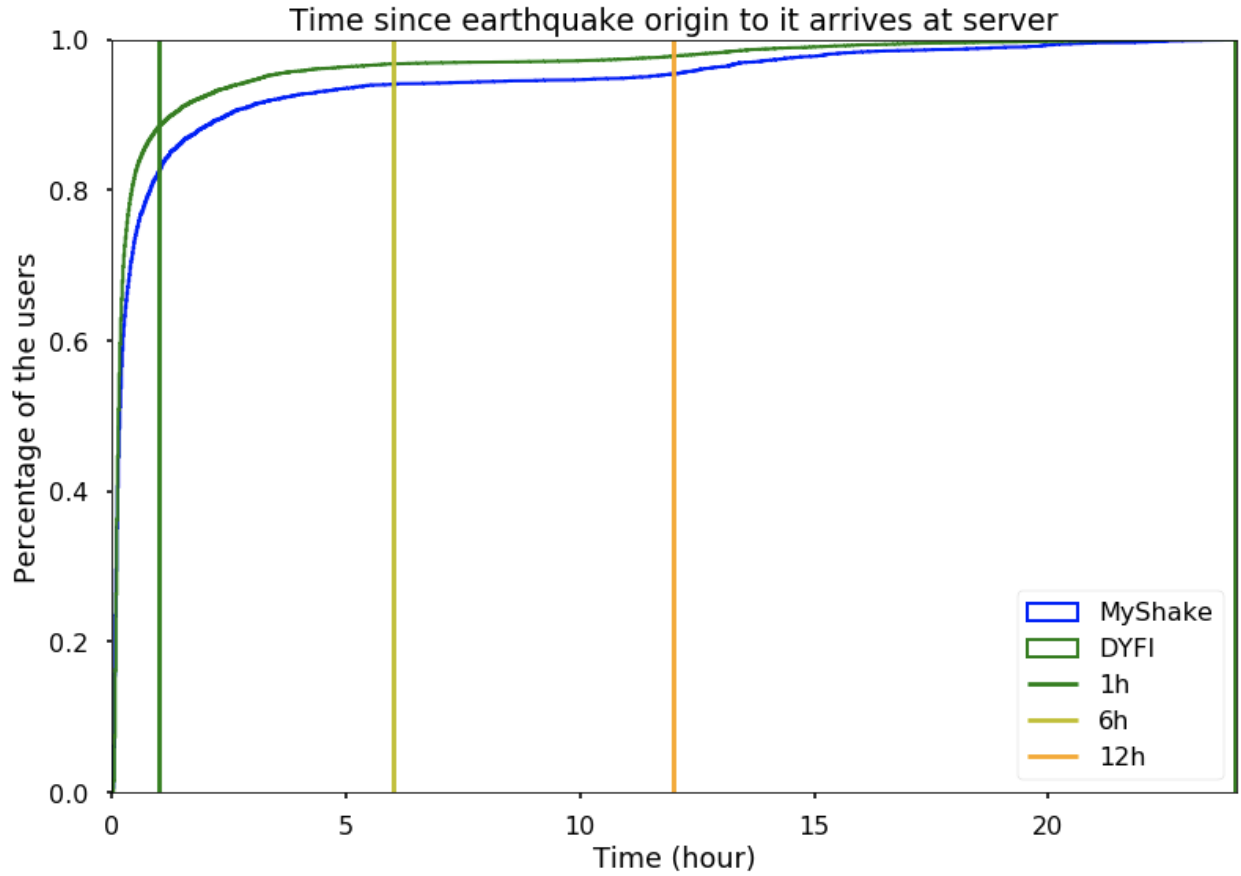

Figure 16. The percentage of the felt reports and DYFI submissions versus time after the origin of the earthquake (nc73512355).